\documentclass[a4paper, twocolumn]{article}

\usepackage{authblk}
\usepackage{graphicx}
\usepackage{dcolumn} 
\usepackage{bm} 
\usepackage{amsmath}
\usepackage{amssymb, graphicx}
\usepackage{hyperref}
\usepackage{physics}
\usepackage{siunitx}
\usepackage{xcolor}
\usepackage{microtype}

\usepackage[style=numeric, sorting=none, hyperref,
    url=false,
	doi=false,
	isbn=false,
	eprint=false,
	giveninits=true,
            sortcites=true,
            sorting=none,
            language=british]{biblatex}
\newbibmacro{string+doi}[1]{%
  \iffieldundef{doi}{%
    \iffieldundef{url}{#1}{\href{\thefield{url}}{#1}}
    }{%
    \href{http://dx.doi.org/\thefield{doi}}{#1}%
  }}
\DeclareFieldFormat{title}{\usebibmacro{string+doi}{\mkbibemph{#1}}}
\DeclareFieldFormat[article]{title}{\usebibmacro{string+doi}{\mkbibquote{#1}}}
\DeclareFieldFormat[thesis]{title}{\usebibmacro{string+doi}{\mkbibquote{#1}}}

\definecolor{lapis}{HTML}{22577A}
\definecolor{cosmos}{HTML}{650D1B}
\definecolor{orange}{HTML}{FF6B35}
\definecolor{asparagus}{HTML}{679436}
\definecolor{grey}{HTML}{4A5759}

\definecolor{pink}{HTML}{D73665}

\newcommand{\cw}{CW}
\newcommand{\Rb}{Rb}

\newcommand{\esRb}{\textsuperscript{87}Rb}

\newcommand{\ladder}{$5S_{1/2} \rightarrow{} 5P_{3/2} \rightarrow{} 5D_{5/2}$}
\newcommand{\ladderone}{$5S_{1/2} \rightarrow{} 5P_{3/2}$}
\newcommand{\laddertwo}{$5P_{3/2} \rightarrow{} 5D_{5/2}$}
\newcommand{\telecommsladder}{$5S_{1/2} \rightarrow{} 5P_{3/2} \rightarrow{} 4D_{5/2}$}
\newcommand{\telecommsladdertwo}{$5P_{3/2} \rightarrow{} 4D_{5/2}$}

\newcommand{\myfigref}[1]{Fig.~\ref{#1}}
\newcommand{\myfigureref}[1]{Figure~\ref{#1}}
\newcommand{\mysubfigref}[2]{Fig.~\hyperref[#1]{\ref*{#1}~(#2)}}
\newcommand{\mysubfigureref}[2]{Subfigure~\hyperref[#1]{\ref*{#1}~(#2)}}

\newcommand{\inlineref}[1]{Ref.~\cite{#1}}

\author[1]{W. O. C. Davis}
\author[2]{P. Burdekin}
\author[1]{T. Wasawo}
\author[2]{S. E. Thomas}
\author[1,3]{P. J. Mosley}
\author[1,3]{J. Nunn}
\author[1,4,*]{C. McGarry}

\affil[1]{Centre for Photonics and Photonic Materials, Department of Physics, University of Bath, Bath, BA2 7AY, UK}
\affil[2]{QOLS, Department of Physics, Imperial College London, London, SW7 2BW, UK}
\affil[3]{ORCA Computing Ltd. 30 Eastbourne Terrace, London W2 6LA UK}
\affil[4]{Currently with School of Physics, University of Sydney, Sydney, New South Wales, Australia
}
\affil[*]{cameron.mcgarry@sydney.edu.au}

\begin{document}

\title{\vspace{-3cm} Fast, low-loss, all-optical phase modulation in warm rubidium vapour}

\maketitle

\textbf{
Low-loss high-speed switches are an integral component of future photonic quantum technologies, with applications
in state generation, multiplexing, and the implementation of quantum gates.
Phase modulation is one method of achieving this switching, but existing optical phase modulators either achieve high bandwidth or low loss, but not both.
We demonstrate fast (\SI{100}{\mega\hertz} bandwidth), low-loss (\SI{83+-2}{\percent} transmission) phase shifting ($\Delta\phi = (\SI{0.90+-0.05}{})\pi$) in a signal field, induced by a control field, and mediated by the two-photon \ladder{} transition in \esRb{} vapour.
We discuss routes to enhance both performance and scalability for application to a range of quantum and classical technologies.
}

\section{Introduction}

Photonics has revolutionised telecommunications since the development of fibre optics in the 1980s.
Photonic data buses are supplanting electronics in high performance computing~\cite{biberman2012optical}, and more recently photonic platforms for machine learning are emerging~\cite{miscuglio2020photonic}.
Looking forwards, photonics can provide a platform for communication with enhanced security by quantum key distribution~\cite{Gisin2007} and support the transfer of quantum information between nodes~\cite{Kimble2008}.
Photonics also represents a promising architecture for the implementation of universal quantum computing~\cite{slussarenko2019photonic}, where engineered non-classical optical states are used to solve computational problems that are intractable with classical (i.e.\ non-quantum) resources.

All of these applications require high-speed switching which can be achieved by phase modulation of an optical signal.
Existing technology offers either a low-loss or high-bandwidth solution, but not both at once.
For example, fibre-integrated electro-optical modulators are commercially mature, and can offer phase modulation on nanosecond timescales.
Nevertheless the insertion losses of these devices
adds a practical overhead: mitigating these losses requires increased input power, intermediate amplifiers, and waste heat management~\cite{Amin}.
Further, increasing demand on switching speeds could lead to the obsolescence of existing semiconductor-based telecommunications devices, driving investigation into all-optical switching technqiues~\cite{Wada2004}.
More efficient technologies for optical modulation are thus desirable across a range of application areas.

Photonic quantum computing represents our primary motivation for this work.
This platform is appealing for a number of reasons, including room-temperature operation of all or many components, high clock-rates, high connectivity, insensitivity to stray fields and modular construction.
But a key technical challenge remains: the requirement to switch and dynamically re-route photons with high speed and extremely low loss.
This is an essential stage in a variety of processes for photonic quantum computing, such as implementing: loop memories~\cite{Pittman2002, Makino2016}, synchronisation~\cite{Kaneda2017} or multiplexing~\cite{Collins2013, Mendoza2016, Francis-Jones2016} of single photon sources and demultiplexing for graph state generation~\cite{Li2020}.
Amplification destroys quantum coherence and so cannot be used to mitigate losses in a quantum system.
The lifetime of photons in a waveguide is limited, hence high bandwidth is required for scalability.
For these reasons, quantum systems have extremely stringent tolerances for speed and loss \cite{pankovich2023high}, which motivates an exploration of alternative platforms that could ultimately deliver better performance than current fibre-integrated electro-optic modulators.

In this article, we describe and demonstrate efficient, all-optical phase modulation of light.
Ideal implementation of phase modulation (a phase shift of $\pi$ radians with no loss) would be equivalent to implementation of an optical switch, by embedding the phase modulator in one arm of a Mach-Zehnder interferometer.
Additionally, we propose that this scheme is compatible with optical fibre systems, by means of fibre-integrated vapour cells, such as those discussed in references~\cite{Haupl2022, Suslov2023, McGarry2024}.


\section{All-optical phase modulation mediated by atoms}

\begin{figure}[t]
    \centering
    \includegraphics[width=0.5\textwidth]{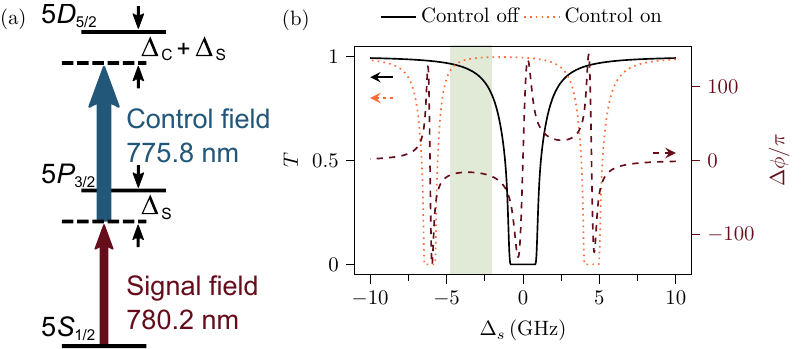}
    \caption{%
    (a) Ladder scheme used for phase modulation of weak signal field (red) by presence of strong control field (blue) detuned from \ladderone{} and \laddertwo{} transitions by $\Delta_s$ and $\Delta_c$ respectively.
    (b) Transmission of the signal field with control off (black, solid) and on (orange, dotted) and relative phase shift of the signal for both control on and control off (red, dashed) through \esRb{} vapour for various $\Delta_s$, as determined by the theoretical model described in the main text.
    Control detuning is fixed at $\Delta_c=\SI{-1.6}{\giga\hertz}$.
    We identify the experimental region of interest (shaded), where there is low-loss of the signal for both control on and control off, as well as a high phase shift.
    Note the two-photon absorption feature at $\Delta_s=\SI{-6}{\giga\hertz}$, which is away from $\Delta_s=-\Delta_c$ due to the a.c.\ Stark-shift.
    }
    \label{theory}
\end{figure}

Our scheme, depicted in \mysubfigref{theory}{a}, takes advantage of the \ladder{} two-photon transition in \esRb{}.
This is the same transition used in electromagnetically-induced transparency (EIT) optical control~\cite{Hendrickson2010, Jones2015, Finkelstein2023, Briscoe2023} and memory~\cite{Finkelstein2018} schemes.
A weak signal field, detuned from resonance with the \ladderone{} transition by frequency $\Delta_s$, counter-propagates through a \esRb{} vapour cell with a high-intensity, pulsed control field.
This control field is detuned from resonance with the \laddertwo{} transition by frequency $\Delta_c$, and its presence induces a change in the susceptibility ($\chi$) of the \esRb{}, as experienced by the signal.
Hence, the control field modulates the phase of the signal field as it traverses the vapour cell.

The susceptibility determines the phase accumulated by light as it traverses the vapour cell, as well as the absorption by the vapour.
We require both sufficiently large change in phase ($\Delta\phi$ must be an odd integer multiple of $\pi$), as well as high transmission $T\sim1$ for the scheme to be viable.
The switching speed is limited to the speed at which the control intensity can be modulated, which can in principle be in the picosecond or even femtosecond regime by using commercially available mode-locked pulsed laser systems.
Using the theoretical results described in \inlineref{GeaBanacloche1995}
we can write the susceptibility as a function of the detunings and the control power ($P_c$), i.e.\ $\chi=\chi(\Delta_s, \Delta_c, P_c)$, to determine the transmission when the control field is on and off, as well as the relative change in phase when the control is turned on.
This calculation is detailed in Code File 1 (\inlineref{Jupyter}), further simulations are described in \inlineref{Siddons}.

Using the \cw{} model, it is possible to demonstrate the existence of regions with high transmission and a large phase shift, as depicted in \mysubfigref{theory}{b} 
(grey shading, $\Delta_s=\SIrange{-4.8}{-2}{\giga\hertz}$ $T=0.96$, $\Delta\phi=-15\pi$) and \inlineref{Jupyter}.
However, we emphasise that the \cw{} model described here is not entirely representative of our experiment or of the applications being considered, where pulsed lasers are used to achieve high control power.
In this case, full time-resolved solutions of the optical Bloch equations are required to accurately describe the system~\cite{seefc}
%
, as will be presented below.

\section{Phase modulation of a \cw{} signal field}
\label{CW}

\begin{figure*}[ptbh]
    \centering
    \includegraphics[width=120mm]{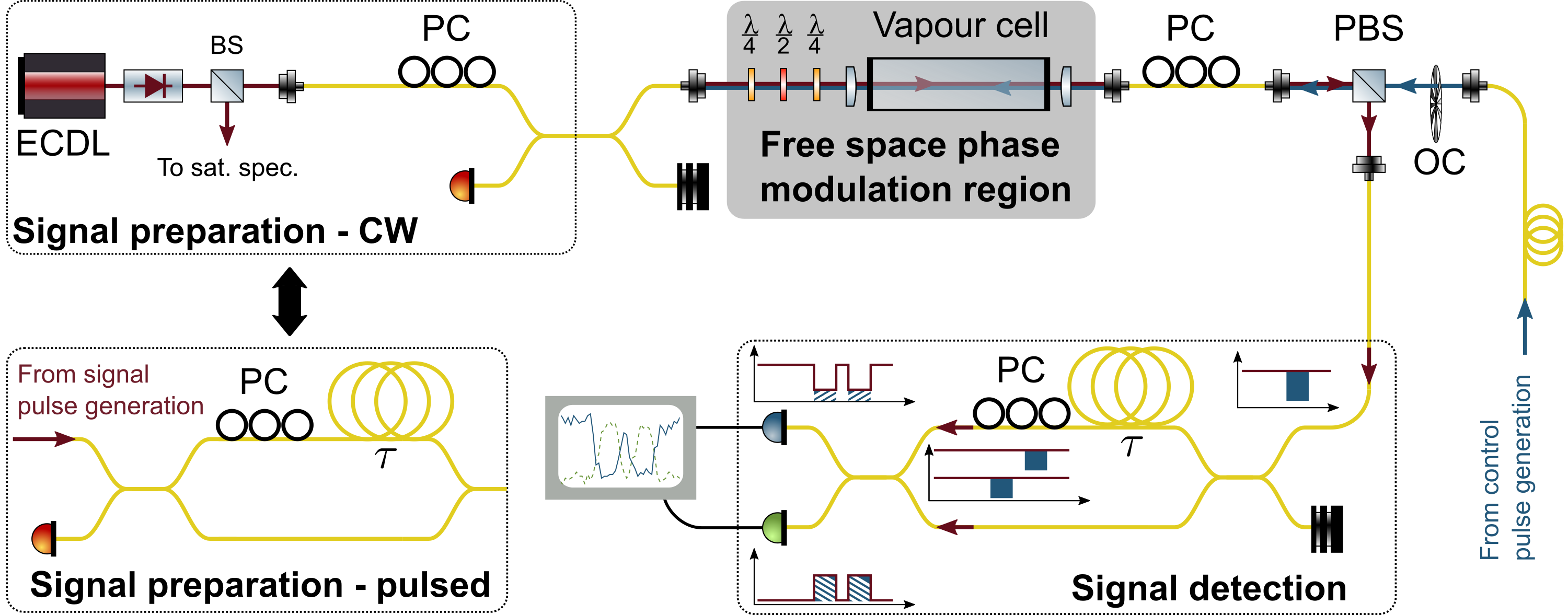}
    \caption{Experimental setup, showing control and signal fields counter-propagating through \esRb{} vapour cell in free space, the generation scheme for pulses and \cw{} signal field, and detection region.
    In the \cw{} case, a beam splitter (BS) is used to calibrate the external cavity diode laser (ECDL) by saturation absorption spectroscopy of \Rb{} and in the pulsed case an optical chopper (OC) is inserted to modulate the control pulses.
    The pulse generation is explained in section~\ref{supp}.
    Polarisation controllers (PC) and waveplates ($\lambda/2$, $\lambda/4$) ensure that the polarisation of the fields is $\sigma^+$ circularly polarised at the vapour cell, and separates the fields at a polarising beam splitter (PBS) for phase-sensitive signal detection by a Franson interferometer with delay $\tau$ in one arm.
    Cartoon graphs in the signal detection region depict operation of the time-binned interferometer for \cw{} signal field.
    Regions of the signal overlapping with a control pulse in the cell are shaded blue. Blue hashing indicates regions of interference at the output. 
    }
    \label{experiment}
\end{figure*}

First we discuss the experimental implementation in which a \cw{} signal field is modulated by a pulsed control field.
This allows the demonstration of a phase shift that varies with the signal detuning $\Delta_s$ for a fixed control detuning $\Delta_c$.
The experiment is shown in \myfigref{experiment}, and uses the \cw{} signal preparation in the upper dashed box.
The \cw{} signal is generated by an external cavity diode laser (MOGLabs CEL) whose frequency is scanned across the region of interest shown in \myfigref{theory}.
Control pulses at \SI{775.8}{\nano\meter} are generated by pulse carving, amplifying and frequency doubling a pulsed telecomms laser operating near \SI{1552}{\nano\meter}.
A full explanation of the pulse generation is available in section~\ref{supp}.
The control pulses are square, with \SI{4}{\nano\second} duration and \SI{47}{\nano\joule} pulse energy.
The signal field average power is \SI{1.6}{\milli\watt}.
Since the timescale of the pulses is much shorter than that of the signal frequency scan, we assume that $\Delta_s$ is effectively constant for each control pulse.

The signal field counter propagates with control pulses through the \SI{7}{\centi\meter} long vapour cell (Precision Glassblowing), which is enriched with the \esRb{} isotope and heated to \SI{97.6}{\celsius}. 
Half- and quarter-waveplates are set so that the light is circularly polarised, maximising the strength of the interaction ~\cite{Olson2006}.
Following this, the signal field is separated into the detection region, which consists of a Franson interferometer~\cite{Franson1989} (see signal detection region of \myfigref{experiment}) with polarisation control used to maximise the fringes.
The interferometer has a $\tau=\SI{4.8}{\nano\second}$ delay line in the upper arm, which allows for interference between regions of the signal field which has experienced the control field (blue shading in \myfigref{experiment}) with regions that have not.
This results in two interference peaks per control pulse, as can be seen in the left side of the figure.
The measured interference in these time bins can indicate the relative phase that the signal acquired due to the presence of the control field.

\begin{figure*}[ptbh]
    \centering
    \includegraphics[width=13cm]{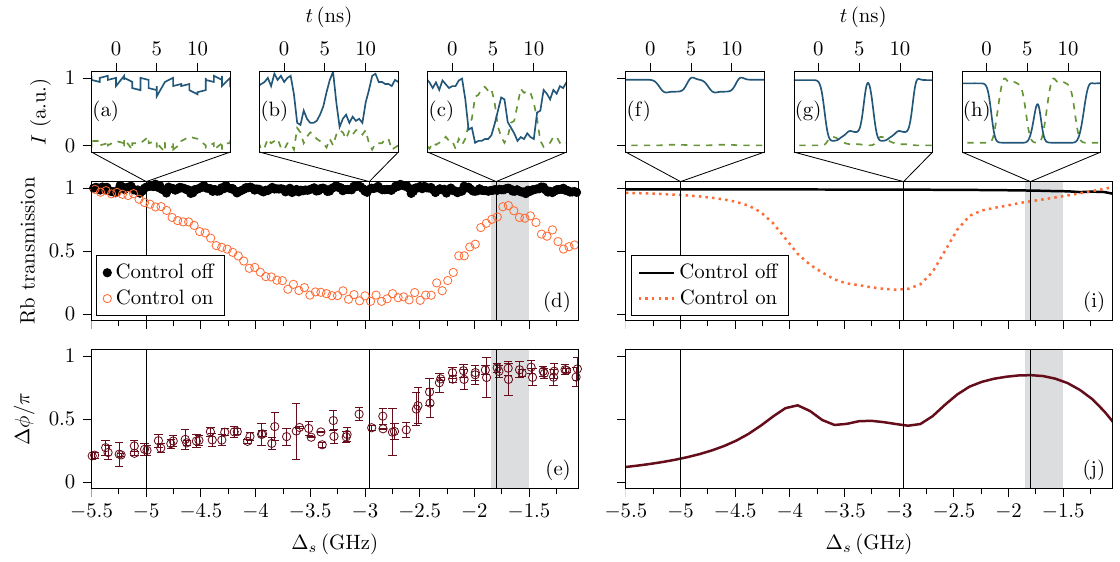}
  \caption{
  Results of \cw{} phase modulation from experiment and simulation are shown in the left and right columns respectively.
  The upper six plots (a, b, c -- experiment; f, g, h -- simulation) show the signal arriving at detectors 1 (green, dashed) and 2 (blue, solid) for various values of $\Delta_s$.
  In each of these traces, $t=0$ corresponds to the detection time of the control pulse.
  Analysis of this interference determines the transmission through the vapour cell (middle panels: d -- experiment; i --- simulation) for the control on (empty orange circles; dashed orange line) and off (filled black circles; solid black line).
  The phase shift is extracted from the contrast of the fringes (bottom panels: e; j).
  }
  \label{cwresults}
\end{figure*}

\myfigureref{cwresults} shows the results of the \cw{} signal field phase modulation for control field fixed at $\Delta_c/2\pi = \SI{1.6}{\giga\hertz}$.
The top row shows a selection of the interference peaks on each detector for various $\Delta_s$.
From these we extract the transmission of the signal through the vapour cell ($T$) both with the control on and the control off, shown in the second row.
This procedure is detailed in section~\ref{supp}.
The third row shows the phase shift between the control on and control off.
Additionally we show theoretical predictions for both transmission and phase modulation (solid lines).
These are given by the aforementioned time dependent solutions to the optical Bloch equations, which are solved for a signal field consisting of one photon.

We observe that for $\Delta_s$ close to zero there is increased absorption due to the single photon resonance, which is broadened by the bright signal field.
We see also that there is significant absorption from the two-photon effect around $\Delta_s=\SI{-3}{\giga\hertz}$.
There is some deviation of theory from experiment approaching the single photon resonance, which we attribute to the discrepancy between the simulated single-signal photon regime, and the bright signal light used in the experiment. In the latter case, the a.c.\ Stark shift of the signal light causes additional broadening, which is commensurate with the effect seen in \inlineref{Pack2007}.
Population transfer due to optical pumping may also be a contributing factor to the difference between experimental and theoretical results seen here.

We identify a signal detuning range of $\Delta_s=\SIrange{-1.85}{-1.5}{\giga\hertz}$ (grey shading in the figure) where the transmission is highest and the phase shift is relatively flat.
We achieve a phase shift $\Delta\phi/\pi=\SI{0.90+-0.05}{}$ with end-to-end transmission $T=\SI{83+-2}{\percent}$, which includes the insertion loss through the vapour cell of \SI{4.5}{\percent}.
Note that this region is consistent for bright signals and single-photon level signals, an therefore useful for classical and quantum photonics.

\section{Phase modulating a pulsed signal field}

In this section we discuss the phase modulation of a $\sim 100$~MHz-bandwidth pulsed signal field, as required for clocked photonic systems.
Signal pulses are generated by the same means as for control pulses, as was described in the previous section. 
Signal field pulses are split, with a time delay of  $\tau=\SI{5}{\nano\second}$, as shown in the lower dashed box of \myfigref{experiment}.
The delay line is used to ensure that both signal and control pulses arrive at the cell together, such that the control overlaps with only the second pulse.
The control beam is chopped at \SI{1}{\kilo\hertz} to allow extraction of the control-induced phase shift on a time-scale fast compared to the interferometer drift.
The optical amplifier gain is now divided between signal and control fields,
which reduces the available control power and consequently the achievable phase shift.

We set $\Delta_c=\SI{2}{\giga\hertz}$, and ramp $\Delta_s$ over a range \SIrange{-2}{0}{\giga\hertz} at a frequency of \SI{1}{\hertz}, much slower than the chopper frequency.
Ramping $\Delta_s$ allows the single-photon absorption line to be resolved when the control is off, so that the stability of the laser can be monitored, and $\Delta_s$ can be determined precisely (although we present results for fixed $\Delta_s$ here).
Any signal pulses that experience a partially chopped control are discarded from the analysis.
As the chopper frequency is three orders of magnitude larger than the signal ramp frequency, the signal laser frequency is roughly constant across the control pulses contained within a single window of the chopper.

\begin{figure}[t]
    \centering
    \includegraphics[width=0.5\textwidth]{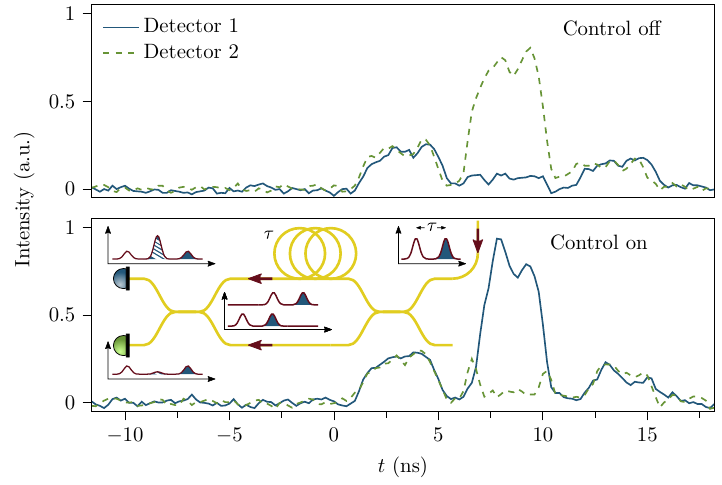}
    \caption{
    Demonstration of phase modulation for pulsed control and signal.
    We show the interferometer output for control off (top panel) and on (bottom panel) for signal pulses as described in the main text.
    Here we set $t=0$ as the time that the first pulse arrives.
    Note the change in the central peak between the control on and off, corresponding to $\Delta\phi/\pi=\SI{0.53+-0.06}{}$
    The inset shows how we interpret the three peaks arriving in at the signal detection arm shown in \myfigref{experiment}, with polarisation controllers omitted for clarity.
    A pulse which has encountered a control pulse in the vapour cell is indicated by blue shading.
    Interference on the central peak at the output is indicated by blue-white hashing, and is used to extract transmission and phase information as in the \cw{} case.
    }
    \label{pulsedresults}
\end{figure}

The results for modulation of the pulsed signal field are shown in \myfigref{pulsedresults}.
On the detectors we see the region of interest from \SI{6}{\nano\second} to \SI{12}{\nano\second}, which corresponds to interference between the pulse that overlapped with the control and the pulse that didn't. Hence the interference depends on the relative phase change induced by the control.
The cell is again at \SI{97.6}{\celsius}, with control pulse energy of \SI{33}{\nano\joule}, signal pulse energy of \SI{200}{\pico\joule}, $\Delta_c=\SI{2}{\giga\hertz}$, $\Delta_s=-\SI{1.9}{\giga\hertz}$, and pulse duration \SI{4}{\nano\second}.
The switch of the interferometer output from detector 2 to detector 1 is due to the presence of the control field, and is a clear indicator of phase modulation.
Comparison of the two right hand peaks shows that the control pulses are changing the shape and magnitude of the signal pulse, due to a combination of absorption and dispersion induced by the control.

The average transmission and phase shift across these pulses is calculated by integration.
This yields a transmission of $T=\SI{84+-6}{\percent}$ for the control field on, and phase shift of $\Delta\phi/\pi = \SI{0.53+-0.06}{}$.
As the control pulse is the same duration as the signal pulse, then the signal pulse does not experience a uniform control intensity distribution in time.
The incomplete temporal overlap reduces the phase modulation effect, as can be seen in the edges of the the detector 2 peak when the control pulse is on in \myfigref{pulsedresults}.

\section{Conclusion}

We have demonstrated low-loss, all-optical phase modulation operating on the timescale of nanoseconds.
Our scheme utilises the change in susceptibility of warm \esRb{} vapour in the presence of a strong control field, inducing a phase shift in a weak signal field.
We are able to demonstrate phase shifts close to $\Delta\phi = \pi$, the critical value required for optical switching.
Our theoretical model predicts that reaching $|\Delta\phi|>10\pi$ is possible with low loss, and that simultaneously increasing the temperature, control intensity and control detuning is predicted to further improve the performance~\cite{Lahad2017}.
Various parameters can be tuned, especially the signal and control detunings; optical pumping can be used to control hyperfine state occupations~\cite{Finkelstein2018}.

We propose that improvement to phase shift and transmission can be made by implementing the phase modulation scheme inside a hollow core optical fibre~\cite{Yu2016} where the core is filled with \esRb{} vapour to realise an in-fibre vapour cell~\cite{Perrella2018, Slepkov2008, Sprague2014, Suslov2021, Suslov2023}.
This would allow for significantly higher intensity, and hence higher phase shift, of the control field, due to the small ($\sim\SI{1000}{\micro\meter\squared}$) size of the mode which can be maintained over lengths of optical fibre much greater than the Rayleigh range.
Such a device would have potential for scalability, due to ability to easily integrate with other optical fibre systems~\cite{Suslov2021}.
An alternative route to enhancement is by implementation in an optical cavity, where the interaction of the control field is increased by a factor of the cavity finesse~\cite{Gorshkov2007, OShea2013, Nunn2017}.
This is particularly appealing for microresonators where high quality factors ($Q>10^6$) are readily achievable~\cite{DelHaye2007, Lee2013, Sumetsky2010, Pfeifer2022, Svela2020}.

Finally, we suggest that phase modulation at other wavelengths may be possible by applying our same scheme at other wavelengths, particularly those that would be compatible with existing telecommunication infrastructure.
Of particular interest may be the \telecommsladder{} transition in \esRb{}, which has recently been used to demonstrate an off-resonant cascaded absorption memory for classical light~\cite{Thomas2023PRApp} and single photons \cite{Thomas2024}.
Here, the signal field is on the \telecommsladdertwo{} transition at \SI{1529}{\nano\meter}, which is close to the telecommunication C-band.
In this scheme, where the \ladderone{} transition is used as the control, weaker control field is required, further reducing overheads.
The phase modulation technique detalled here offers a route towards waveguide-integrated, low-loss, high-speed and all optical switching operating at a range of wavelengths.
This presents the potential for significant impact for both classical and quantum photonics.

\section{Supplementary Material}
\label{supp}

\subsection{Pulse generation} \label{Pulse generation}
 
\begin{figure}[hbt!]
    \centering
    \includegraphics[width=0.4\textwidth]{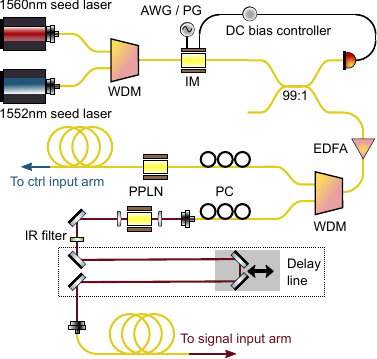}
    \caption{Setup used to generate signal and control pulses. SMF, single-mode fibre; PMF, polarisation-maintaining fibre; FPC, fibre polarisation controller; AWG, arbitary waveform generator; IM, intensity modulator; PD, photodiode; PPLN, periodically poled lithium niobate; WDM, wavelength division multiplexer; EDFA, erbium doped fibre amplifier.}
     \label{pulses}
\end{figure}

The experimental setup starts with generating two trains of pulses.
Strong control pulses at \SI{776}{\nano\meter} and weak signal pulses at \SI{780.2}{\nano\meter}.
The pulse generation scheme is shown in Fig.~\ref{pulses}. Two tunable fibre coupled seed lasers in the telecomms C band at \SI{1552}{\nano\meter} (control) and \SI{1560}{\nano\meter} (signal) (ID Photonics CoBrite DX1, and Eblana EP1560-0-DM-B05-FM, respectively) are used as seed lasers. Both wavelengths are combined using a wavelength division multiplexer (Oz Optics WDM-12N-111-1552/1560-8/125-PPP-60-3A3A3A-1-1).
The seed lasers are pulsed carved using a waveguided Lithium Niobate intensity modulator (EOSpace - AZ-DS5-10-PFA-PFA-LV), stabilised by a DC bias controller (MBC-HER-PD-3A-0V) at the operating point for minimum transmission. Fast electrical pulses are applied by a radio-frequency (RF) pulse generator (Picoquant PPG512) with the repetition rate controlled by an arbitrary waveform generator.
This allows a stable train of signal and control pulses to be carved, and the pulse duration can be varied between \SI{0.6}{\nano\second} and \SI{4}{\nano\second}.

The signal and control pulses are amplified in an Erbium doped fibre amplifier (EDFA). The amplified pulses are then split out using a second wavelength division multiplexer (Oz Optics WDM-12N-111-1552/1560-8/125-SSS-60-3A3A3A-1-1). The 1552\,\si{\nano\meter} control field passes into a waveguided periodically poled lithium niobate crystal (Covesion WGP-H-1552-40) for frequency doubling to 776\,\si{\nano\meter}. The 1560\,\si{\nano\meter} signal field is coupled into free-space and is focused into a bulk periodically poled lithium niobate crystal (Covesion MSHG1550-1.0-40) for frequency doubling to 780.2\,\si{\nano\meter}. Any remaining 1560\,\si{\nano\meter} light is removed by a short-pass filter. The 780\,\si{\nano\meter} signal light passes through a variable optical delay line before being coupled back into fibre. The optical delay line allows the relative delay between the signal and control fields to be adjusted to ensure that pulses meet inside the Rb vapour cell. 

\subsection{Pulse analysis}

We now describe the analysis of the pulses that arrive at the photodetectors after the vapour cell and Franson interferometer (see Fig.~2 of the main text). This procedure is used to convert the interferometer data (see main text Fig.~4~(a-c)) into the phase shift and transmission due to the rubidium vapour cell in the case of a continuous wave (c.w.) signal field.

The frequency of the scanning signal field can be calibrated by summing the channels and applying a band-pass filter to the voltage on either photodiode. This removes the high-frequency effect of the interferometer and the modulation due to the control pulses. The single photon feature can then be identified, and used as a reference.

The filter is then removed to observe the effects of the control field and the interferometer.
The pulse train arriving at the detectors is observed. In the c.w.\ signal field case, this will result in pairs of pulses arriving at the detectors, as described in the main text.
When the control field is off, we observe interference due to the Franson interferometer alone, and the free spectral range is measured to be \SI{200}{\mega\hertz}.
The positions of interferometer peaks are found using the Scipy Python package~\cite{python, scipy}.

The control field is then turned on, so that pulses counter-propagate through the vapour cell, against the signal field.
The photodiode in the signal preparation arm is used to detect these control pulses, which we see are synchronised with interference in the signal detection arm.
The contrast of these fringes can then be analysed to determine the induced phase modulation as follows~\cite{Franson1989, WDThesis}.

\begin{figure}[hbt!]
    \centering
    \includegraphics[width=0.4\textwidth]{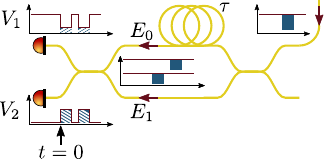}
    \caption{
    Franson interferometer, highlighting field in each arm before the second fibre coupler ($E_{0,1}$) and the voltages at each photodiode $V_{1,2}$.
    }
     \label{interferometer}
\end{figure}

Consider the Franson interferometer used for signal detection, as shown in Fig.~\ref{interferometer}.
The time delay on the upper arm, $\tau$, is assumed to be much longer than the spacing between control pulses, but short on the timescale of the signal frequency scan, so that we can consider all light passing through the interferometer to be of the same frequency.
By synchronisation with the control detection photodiode, we are able to analyse the change in interference due to the control pulse (denoted by the blue shading).

The electric field that has not experienced a control pulse has value just before the second fibre coupler of $E_0$.
The other arm has a field $E_1 = E_0te^{-i(\Delta\phi + kc\tau)}$. 
Note that by symmetry, we can assume the phase-modulated light to always be in the lower arm, however a limitation of this setup is that only $\cos(\Delta\phi)$ can be determined here, and hence the sign of the phase shift is unavailable (we take it to be positive in this work).

We have introduced $t=e^{-\alpha L /2}$ as the coefficient of transmission through the vapour cell, $L$ is the cell length and $alpha$ is the absorption coefficient.
The phase shift due to the rubidium is $\Delta \phi$, and the second term in the phase exponent ($kc\tau$) is interference due to the interferometer delay ($k$ is the signal field wavenumber and $c$ is the speed of light).
The voltage output off the two photodiodes is given by
\begin{align}
V_1 &\propto |E_0 + \gamma E_1|^2 \\
V_1 &= a\left(1 + t^2 + 2\gamma t\cos(\Delta\phi + kc\tau)\right) \\
V_2 &= a\left(1 + t^2 - 2\gamma t\cos(\Delta\phi + kc\tau)\right)
\end{align}
where $a$ is a constant of proportionality and $\gamma$ has been introduced to represent mismatch in the polarisation of the two arms.
This latter parameter is complex and $|\gamma|=1$.
We will see that $\gamma$ is determined in our analysis, however practically we are able to control this parameter via the polarisation controller in the interferometer, and can take it to be $1$ (i.e.~the polarisations are complete matched).

The interferometer can be calibrated by considering the voltage at each detector when the control filed is off ($V_i^\text{(ctrl off)}$).
In this situation, we have $\Delta\phi=\alpha=0$, and so we can define
\begin{equation}
a = \frac{V_1^\text{(ctrl off)} + V_2^\text{(ctrl off)}}{4}
\end{equation}
Similarly, we compare the contast of the interference fringes to determine that 
\begin{equation}
\gamma = \frac{V_1^\text{(ctrl off, max)} - V_2^\text{(ctrl off, min)}}{4a}, 
\end{equation}
where we have introduced the maximum ($V_i^\text{(ctrl off, max)}$) and minimum ($V_i^\text{(ctrl off, min)}$) voltages across the signal field scan.

When the control field is turned on, the interferometer phase can be determined using the above constants, and the photodiode signal immediately before the control pulse arrives
\begin{equation}
    \cos(kc\tau) = \frac{V_1 - V_2}{4\gamma a}.
\end{equation}

When the control pulse arrives, $alpha$ and $\Delta \phi$ can be found by reading off $V_1$ and $V_2$ and solving equations (S2) and (S3) numerically.
This is possible since all constants (that is $a$, $\gamma$ and $\cos(kc\tau)$) have been found by the above methodology, and hence this problem is reduced to solving two equations with two unknown values.

\section*{Acknowledgements}

This work was supported by the UK Hub in Quantum Computing and Simulation, part of the UK National Quantum Technologies Programme with funding from UKRI EPSRC Grant EP/T001062/1.
This work is partially funded by Innovate UK Quantum Data Centre of the Future grant 10004793.


\printbibliography

\end{document}